\documentstyle[seceq,preprint,wrapft,epsf]{ptptex}

\title{Many-Particle and Many-Hole States in Neutron-Rich Ne Isotopes
Related to Broken N=20 Shell Closure}

\author{
Masaaki {\sc Kimura}$^{1}$
and Hisashi {\sc Horiuchi}$^{2}$
}

\inst{
$^1$RI-beam Science Laboratory, RIKEN ({\it The Institute of
Physical and Chemical Research}), Wako, Saitama 351-0198, Japan.\\
$^2$Department of Physics, Kyoto University, Kitashirakawa, Kyoto
606-8502, Japan}

\recdate{
}

\abst{
The low-lying level structures of $^{26}{\rm Ne}$, $^{28}{\rm Ne}$ and
$^{30}{\rm Ne}$ which are related to the breaking of the N=20 shell
closure have been studied by the framework of the deformed-basis
antisymmetrized molecular dynamics plus generator coordinate method
using Gogny D1S force. The properties of the many-particle and many-hole
states are discussed as well as that of the ground band. We predict that  
the negative-parity states in which neutrons are promoted
into $pf$-orbit from $sd$ orbit have the small excitation energy in the
cases of $^{28}{\rm Ne}$ and $^{30}{\rm Ne}$ which, we regard, is a
typical phenomena accompanying the breaking of N=20 shell closure. It is
also found that the neutron $4p4h$ structure of $^{30}{\rm Ne}$ appears
in low excitation energy which contains $\alpha$+$^{16}{\rm O}$
correlations. 
}

\begin{document}

\maketitle

\section{Introduction}\label{sec:introduction}
In these years, many experimental and theoretical efforts to investigate
the properties of nuclei away from the stability valley have shown
the variety of the nuclear binding systems  which is beyond our standard 
understanding of the nuclear properties established in the
study of the stable nuclei. The discovery of the neutron-halo
phenomenon \cite{ref:halo} proved that the basic concept of the
density saturation does not hold near the neutron drip-line. Another
important concept of the shell structure and the magic number is also
under the reconsideration because of their rearrangement in 
neutron-rich nuclei which is deduced from the breaking of the neutron
shell closure in neutron-rich N=20 isotones.  

The breaking of the N=20 shell closure in neutron-rich nuclei was 
firstly pointed out from the observation of the anomalous ground state
spin of $^{31}{\rm Na}$ \cite{ref:thibault} associated with the prolate 
deformation. Theoretically, it was suggested that the large
deformation which is caused by the promotion of the neutrons from the
$sd$-shell to $f$-shell possibly overcomes the N=20 shell effect in the
neutron-rich N=20 isotones such as $^{31}{\rm Na}$ and $^{32}{\rm Mg}$
\cite{ref:campi}. The shell model studies
\cite{ref:watt,ref:poves,ref:brown} also
support the neutron promotion into $pf$-shell and a striking result
was given in $^{32}{\rm Mg}$\cite{ref:otsuka} that neutron
$2p2h$ configuration (two holes in the $sd$ shell and two
particles in the $pf$ shell) dominate the ground state of $^{32}{\rm
Mg}$ and the large B($E2;0^+_1\rightarrow2^+_1$) due to the large
deformation was predicted. Since the first observation of the 
B($E2;0^+_1\rightarrow2^+_1$) value in $^{32}{\rm Mg}$
\cite{ref:motobayashi} which confirmed the shell model prediction, many
experimental and theoretical studies have been devoted to Z$\sim$10
and N$\sim$20 nuclei
\cite{ref:dean,ref:utsuno_1,ref:peru,ref:stoitsov,ref:egido_1,ref:caurier,ref:utsuno_2,ref:kimura,ref:egido_2,ref:Klotz,ref:sakurai,ref:pritychenko,ref:chiste,ref:yoneda,ref:azaiez,ref:mittig}. 
And nowadays, the systematic large 
deformation and breaking of the N=20 shell closure are theoretically
expected in the ground states of  $^{28-32}{\rm Ne}$, $^{29-33}{\rm Na}$
and $^{28-34}{\rm Mg}$ and experimentally investigated in $^{32,34}{\rm 
Mg}$. Recently, a new NN effective interaction which has a stronger
$(\sigma\cdot\sigma)(\tau\cdot\tau)$ term was suggested to  explain the
drastic change of the location of the neutron drip-line from oxygen  
isotopes to fluorine isotopes based on the monte carlo shell model
study \cite{ref:utsuno_2}. They argued that the extension of the
neutron drip-line in F isotopes is caused by the mixing of the neutron
$0p0h$, $2p2h$ and $4p4h$ configurations which is driven not by the
deformation but by the stronger $(\sigma\cdot\sigma)(\tau\cdot\tau)$
interaction.  

However, the experimental and theoretical information of low-lying states
of neutron-rich isotopes is not enough yet. Especially, negative-parity
states of even-even nuclei are not still confirmed in
experiments\cite{ref:Klotz,ref:mittig} and only little theoretical study
has been made\cite{ref:brown}. The neutron $2p2h$ dominance in the
ground states of neutron-rich N=20 isotones implies that one or three
neutrons can be easily promoted into $pf$-shell with small excitation
energy which leads to the existence of the low-lying negative-parity
states. Indeed, the shell model study\cite{ref:brown} has shown the
possible existence of the low-lying neutron $1\hbar\omega$ and   
$3\hbar\omega$ states in this mass region. Furthermore, the parity of
the ground state of even-odd nuclei $^{33}{\rm Mg}$ is identified to be
the positive\cite{ref:nummela,ref:morton} which  means the neutron
$1\hbar\omega$ structure 
of the ground state.  Therefore it is of importance to study
negative-parity  states for understanding the structure of neutron-rich
N$\sim$20 nuclei. In the same sense, the excited positive-parity bands
which will have the neutron $0p0h$ and $4p4h$ structures dominantly are
also  important. Especially, in the case of the $4p4h$ structure, due to
the presence of the four neutrons in the $pf$-shell, the protons which
are almost frozen in the ground state might be activated. Actually, we
will  show that the ($\alpha$+$^{16}{\rm O}$+valence-neutrons type)
cluster-like correlations can exist in the neutron $4p4h$ structure of
$^{30}{\rm Ne}$ which appears in rather low energy region.       

The purposes of this article is  to provide theoretical
information on the properties of the negative-parity states and the
excited bands of the positive-parity of the even-even Ne isotopes to
understand the structure of the neutron-rich $sd$-shell nuclei around
the broken N=20 shell closure. The low-lying level structures of
$^{26}{\rm Ne}$, $^{28}{\rm Ne}$ and $^{30}{\rm Ne}$ are discussed by
using the deformed-basis AMD+GCM framework (deformed-basis
antisymmetrized molecular dynamics plus generator coordinate
method). The deformed-basis AMD 
framework has now been confirmed to describe well the mean-field like
the Hartree-Fock method in addition to the ability of describing well the
cluster structure. Since in AMD the energy variation is made after
parity projection, the AMD can describe negative parity states which
almost all Hartree-Fock calculations cannot treat. The use of AMD
also enables us to study the possible 
existence of the $\alpha$+$^{16}{\rm O}$+valance-neutrons type structure
in neutron-rich Ne isotopes. Since the low-lying states of $^{20}{\rm
Ne}$ have the $\alpha$+$^{16}{\rm O}$ cluster structure, the cluster
correlation may survive even in the neutron-rich isotopes, like the
case of neutron-rich Be isotopes which inherit the cluster structure of
$^{8}{\rm Be}$.  It will be shown that the excitation energy of the
negative-parity states becomes lower around N=20 because of the breaking
of the N=20 shell closure and the $\alpha$+$^{16}{\rm O}$ cluster
correlations appears in the positive-parity $4p4h$ states.

The contents of this article is as follows. In the next section, the
theoretical framework of the deformed-basis AMD is briefly
explained. In the section \ref{sec::result}, the obtained energy
curves and level schemes are discussed. The $E2$ transition
probabilities are also studied. The density distribution and the possible
existence of the cluster core in the ground and excited states of
$^{26}{\rm Ne}$, $^{28}{\rm Ne}$ and $^{30}{\rm Ne}$ are examined.
In the last section, we summarize this work.

\section{Theoretical Framework}\label{sec:framework}
In this section, the framework of the deformed-basis AMD+GCM is explained
briefly. For more detailed explanation of the framework of the
deformed-basis AMD, readers are referred to references
\cite{ref:ono,ref:enyo}. 
The intrinsic wave function of the system with
mass A is given by a Slater determinant of single-particle wave packets
$\varphi_i({\bf r})$; 
\begin{eqnarray}
 \Phi_{int} &=& \frac{1}{\sqrt{A!}}\det
  \{\varphi_1,\varphi_2,...,\varphi_A \} ,\label{EQ_INTRINSIC_WF}\\
 \varphi_i({\bf r}) &=& \phi_i({\bf r})\chi_i\xi_i ,
\end{eqnarray}
where the single-particle wave packet $\varphi_i$ consists of the
spatial $\phi_i$, spin $\chi_i$ and isospin $\xi_i$
parts. Deformed-basis AMD employs the triaxially deformed Gaussian
centered at ${\bf Z}_i$ as the spatial part of the single-particle wave
packet. 
\begin{eqnarray}
 \phi_i({\bf r}) &\propto& \exp\biggl\{-\sum_{\sigma=x,y,z}\nu_\sigma
  (r_\sigma - {\rm Z}_{i\sigma})^2\biggr\},\nonumber\\
 \chi_i &=& \alpha_i\chi_\uparrow + \beta_i\chi_\downarrow,
  \quad |\alpha_i|^2 + |\beta_i|^2 = 1\nonumber \\
 \xi_i &=& proton \quad {\rm or} \quad neutron. \label{EQ_SINGLE_WF}
\end{eqnarray}
Here, the complex number parameter ${\bf Z}_i$ which represents the center
of the Gaussian in the phase space takes independent value for each
nucleon. The width parameters $\nu_x, \nu_y$ and $\nu_z$ are real number
parameters and take independent values for $x$, $y$ and $z$ directions,
but are common to all nucleons. Spin part $\chi_i$ is parametrized by
$\alpha_i$ and $\beta_i$ and isospin part $\xi_i$ is fixed to up
(proton) or down (neutron). ${\bf Z}_i$, $\nu_x, \nu_y, \nu_z$ and
$\alpha_i$, $\beta_i$ are the variational parameters and optimized by
the method of frictional cooling. The advantage of the
triaxially deformable single-particle wave packet is that it makes
possible to describe the cluster-like structure and deformed mean-field
structure within a single framework which discussed in
reference. \cite{ref:enyo}.  

As the variational wave function, we employ the parity projected wave
function as in the same way as many other AMD studies
\begin{eqnarray}
 \Phi^{\pm} = P^\pm \Phi_{int} = \frac{(1\pm P_x)}{2} \Phi_{int} ,\label{EQ_PARITY_WF}
\end{eqnarray}
here $P_x$ is the parity operator and $\Phi_{int}$ is the intrinsic wave
function given in Eq(\ref{EQ_INTRINSIC_WF}). Parity projection  makes
possible to determine the different structure of the intrinsic state for
the different parity states. 

Hamiltonian used in this study is as follows;
\begin{eqnarray}
\hat{H} = \hat{T} + \hat{V_n} + \hat{V_c} - \hat{T_g} ,
\end{eqnarray}
where $\hat{T}$ and $\hat{T}_g$ are the total kinetic energy and the
energy of the 
center-of-mass motion, respectively.
We have used the Gogny force with D1S parameter set as an effective
nuclear force $\hat{V}_n$. Coulomb force $\hat{V}_c$ is approximated by
the sum of seven Gaussians. 
The energy variation is made under a constraint on the nuclear
quadrupole deformation by adding to $\hat{H}$ the constraint potential
$V_{cnst}=v_{cnst}(\langle\beta\rangle^2-\beta_0^2)^2$ with a large
positive value for 
$v_{cnst}$. At the end of the variational calculation, the expectation
value of $V_{cnst}$ should be zero in principle and in the practical
calculation, we confirm it is less than 0.1keV. The optimized wave
function is denoted by $\Phi^{\pm}(\beta_0)$. Here it should be noted
that this constraint does not refer to the deformation parameter
$\gamma$, which means that $\Phi^{\pm}(\beta_0)$ with positive $\beta_0$
can be not only prolate but also oblate. 

From the optimized wave function, we project out the eigenstate of the
total angular momentum $J$,
\begin{eqnarray}
 \Phi^{J\pm}_{MK}(\beta_0) = P^{J}_{MK}\Phi^{\pm}(\beta_0)
  = P^{J\pm}_{MK}\Phi_{int}(\beta_0).
  \label{EQ_ANGULAR_WF}
\end{eqnarray} 
Here $P^{J}_{MK}$ is the total angular momentum projector. The integrals
over the three Euler angles included in the $P^{J}_{MK}$ are evaluated by
the numerical integration. 

Furthermore, we superpose the wave functions $\Phi^{J\pm}_{MK}$ which
have the same parity and the angular momentum but have different value of
deformation parameter $\beta_0$ and $K$. Thus the final wave function of
the system becomes as follows;
\begin{eqnarray}
 \Phi_n^{J\pm} = c_n\Phi^{J\pm}_{MK}(\beta_0)
  + c_n^\prime\Phi^{J\pm}_{MK^\prime}(\beta_0^\prime) + \cdots,
  \label{EQ_GCM_WF}
\end{eqnarray}
where other quantum numbers except total angular momentum and parity
are represented by $n$. The coefficients $c_n$, $c'_n$,... are determined
by the Hill-Wheeler equation,
\begin{eqnarray}
 \delta \bigl(\langle\Phi^{J\pm}_n|\hat{H}|\Phi^{J\pm}_n\rangle - 
  \epsilon_n \langle\Phi^{J\pm}_n|\Phi^{J\pm}_n\rangle \bigr) =0.
  \label{EQ_GCM_EQ}
\end{eqnarray}

\section{Low-lying level structure of neutron-rich {$\bf Ne$} isotopes}\label{sec::result}
We have studied low-lying level schemes of the neutron-rich Ne
isotopes, $^{26}{\rm Ne}$, $^{28}{\rm Ne}$ and $^{30}{\rm Ne}$ to study
the change of the shell structure toward N=20 and the 
mechanism of the nuclear excitation in these isotopes. For the
positive-parity levels, the possible existence of the
$\alpha$+$^{16}{\rm O}$ core is also examined. 

Before the discussion on the obtained results, we explain the analysis
of the obtained wave function and the notations used in this study to
clarify our discussions. In this study, the single-particle structure of
obtained wave function is analyzed by constructing the Hartree-Fock
single-particle Hamiltonian from the obtained AMD wave
function\cite{ref:dote} is as follows. 

When the optimized wave function 
$\Phi^{\pm}=P^{\pm}\frac{1}{\sqrt{A!}}\det \{\varphi_1,\varphi_2,...,\varphi_A
\}$  is given, we calculate the orthonormalized basis $\phi_\alpha$
which is a linear combination of the single-particle wave packets
$\varphi_i$, 
\begin{eqnarray}
 \phi_\alpha=\frac{1}{\sqrt{\mu_\alpha}}\sum_{i=1}^A c_{i\alpha} \varphi_i.
\end{eqnarray}
Here, $\mu_\alpha$ and $c_{i\alpha}$ are the eigenvalue and eigenvector
of the overlap matrix $B_{ij}\equiv\langle\varphi_i|\varphi_j\rangle$,
\begin{eqnarray}
 \sum_{j=1}^{A}B_{ij}c_{j\alpha}=\mu_\alpha c_{i\alpha},
\end{eqnarray}
and it is clear that $\phi_\alpha$ is orthonormalized from this
relation. Using this basis set of $\phi_\alpha$, we calculate the
Hartree-Fock single-particle Hamiltonian $h_{\alpha\beta}$ which is
defined as, 
\begin{eqnarray}
 h_{\alpha\beta} &\equiv& \langle \phi_\alpha|\hat{t}|\phi_\beta\rangle 
+ \sum_{\gamma=1}^A 
\langle\phi_\alpha\phi_{\gamma}|\hat{v}_{Gogny}+\hat{v}_{Coulomb}|
\phi_\beta\phi_\gamma- \phi_\gamma\phi_\beta\rangle \nonumber\\
&&+\frac{1}{2}\sum_{\gamma\delta}\langle\phi_{\gamma}\phi_{\delta}|\phi^*_\alpha\phi_\beta\frac{\partial \hat{v}_{Gogny}}{\partial \rho}|\phi_\delta\phi_\rho-\phi_\rho\phi_\delta\rangle, \label{eq::hf_hamiltonian}
\end{eqnarray}
where $\hat{t}$, $\hat{v}_{Gogny}$ and $\hat{v}_{Coulomb}$ denote the
kinetic operator, the Gogny force, and the Coulomb
potential. ${\partial \hat{v}_{Gogny}}/{\partial \rho}$ denotes the
derivative of the density dependent term of the Gogny force. 

By the diagonalization of $h_{\alpha\beta}$, we obtain the
single-particle energy $\epsilon_s$ and single-particle wave function
$\tilde{\phi}_s$. 
\begin{eqnarray}
  \sum_{\beta=1}^{A}h_{\alpha\beta} f_{\beta s} =  \epsilon_s f_{\alpha s},\\
 \tilde{\phi}_s = \sum_{\alpha=1}^{A} f_{\alpha s}\phi_\alpha. 
\end{eqnarray}
We note that the single-particle energy $\epsilon_s$ and wave function
$\widetilde{\phi}_s$ are obtained 
for occupied states but not for unoccupied states from this
method. Furthermore, since the actual variational calculation is made
after the parity projection (the superposition of the two
Slater determinants), it does not allow the naive interpretation of
$\phi^\pm(\beta_0)$ by the single-particle picture. However, the
single-particle structure obtained by this method is useful to
understand the particle-hole structure of the obtained wave
function. From the parity $\pi_s^\pm=\langle\widetilde{\phi}_s|P^\pm
|\widetilde{\phi}_s\rangle$ and the the angular momentum  
$\langle\widetilde{\phi}_s|\hat{j}_z|\widetilde{\phi}_s\rangle$ in the 
intrinsic frame, we have judged the particle-hole structure. From this
analysis, we find that the obtained wave functions have various
single-particle structure. For example, in the case of the
positive-parity states of $^{30}{\rm Ne}$, neutron $0p0h$, $2p2h$ and
$4p4h$ structures with respect to N=20 appear in the low-lying
states. When protons are excited, proton $2p0h$, $3p1h$ and $4p2h$
structures with respect to Z=8 will be combined with the neutron $ph$
structures. For the convenience, in the following, `${\rm m}p{\rm n}h$
structure' denotes the neutron $ph$ structure. When we need to
distinguish it  from that of protons, it will be written explicitly.

\subsection{Energy curves}
\begin{figure}
\epsfxsize =\hsize
\epsfbox{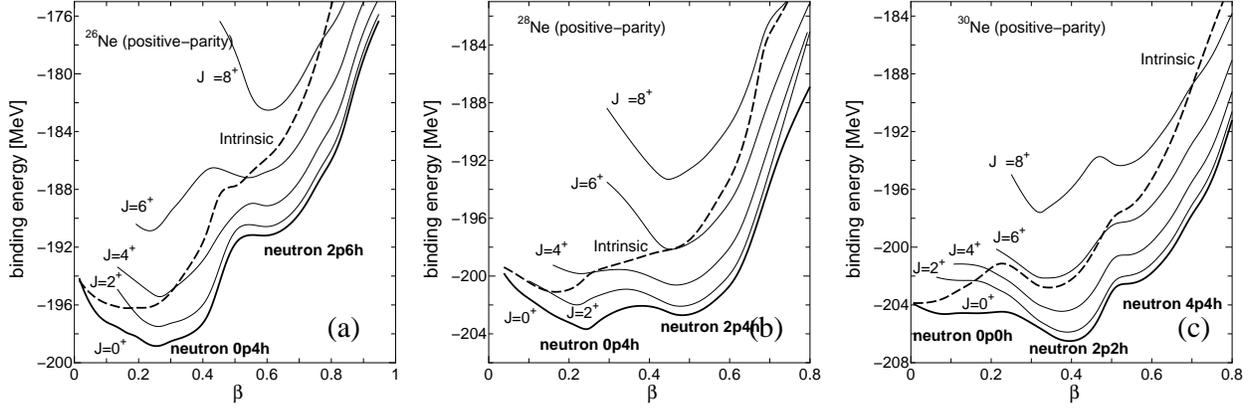}
\caption{Obtained energy curves of the positive parity states of
 $^{26}{\rm Ne}$ (a), $^{28}{\rm Ne}$ (b) and $^{30}{\rm Ne}$ (c). In
 each panel, the energies of the parity projected intrinsic state
 (dashed line), the parity and angular momentum projected states are
 plotted as the functions of the matter quadrupole deformation parameter
 $\beta$.}\label{fig::positive_surface} 
\end{figure}
\begin{figure}
\epsfxsize =\hsize
\epsfbox{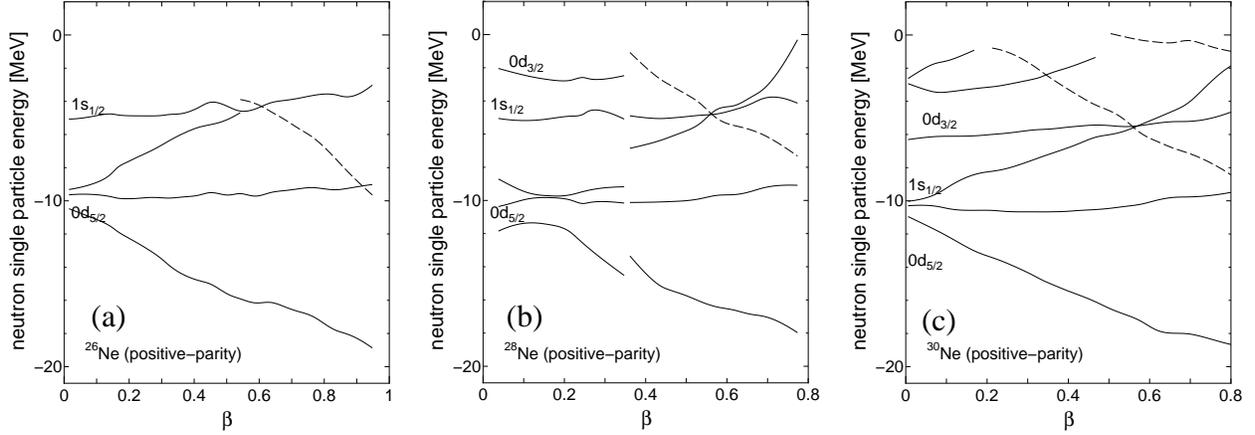}
\caption{The neutron single particle energies of $^{26}{\rm Ne}$ (a),
 $^{28}{\rm Ne}$ (b) and $^{30}{\rm Ne}$ (c) obtained from the
 intrinsic wave functions on the positive-parity energy curves. The
 positive (negative) party single particle orbits are plotted by the
 solid (dashed) lines. Each single particle orbits are occupied by two
 neutrons. }\label{fig::positive_spe}    
\end{figure}

The energy curves as function of the matter quadrupole deformation
parameter $\beta$ obtained before and after the angular momentum
projection are shown in figure \ref{fig::positive_surface} for
positive-parity states. The angular-momentum projected energy curve
of the positive-parity state of $^{30}{\rm Ne}$ (figure 
\ref{fig::positive_surface} (c)) has an energy minimum around
$\beta\sim0.4$ in each spin state. The energy gain of this deformed
state is due to the restoration of the rotational symmetry and it
amounts to about 5MeV in the case of the $J^\pi$=$0^+$ state. This
deformed minimum state has a neutron $2p2h$ structure. The
neutron single-particle energies of $^{30}{\rm Ne}$ are plotted as 
functions of deformation in figure \ref{fig::positive_spe} (c). From
$\beta$=0 to $\beta$=0.2, all neutrons are almost frozen in the N=20
shell closure. This inactivity of neutrons explains the absence of the
$6^+$ and $8^+$ state in this region, since only two protons in
$sd$-shell contribute to the total spin of the system. From $\beta$=0.2
to $\beta$=0.5, two neutrons are promoted into the $pf$-shell
($2p2h$) and the system is most deeply bound. The largely
deformed state ($\beta>$0.5) has a $4p4h$ structure in
which four neutrons are promoted into the $pf$-shell. This state has not
been discussed in detail in other model studies. One of the reason may
be that the stability of the mean-field structure of this state is
doubtful.  Indeed, in our result, this state does not appear as a local
minimum on the energy curve, but as a shoulder around
$\beta$=0.5. However, when we superpose the wave functions on the energy
curve, this state contributes to the $K^\pi$=$0^+_3$ band and is
stabilized through the orthogonalisation against the lower two
bands. The parity 
projection before variation also helps to lower the energy of the $4p4h$
structure. The energy of this state is lowered by about 2MeV by the
parity projection, while $0p0h$ and $2p2h$ structure is not so affected
(less than 1MeV). Note that the $4p4h$ state has a parity asymmetric
intrinsic density distribution.  This asymmetry is the reason why the
energy of the $4p4h$ structure is lowered by the parity
projection. Furthermore, the proton density distribution implies the
possible existence of a $\alpha$+$^{16}{\rm O}$ cluster structure. 
Indeed, the overlap between the proton wave function of this state and
that of the $^{20}{\rm Ne}$ which has an $\alpha$+$^{16}{\rm O}$ cluster
structure is quite large and it amounts to about 80\%. Therefore, the
interpretation of this state by the $\alpha$+$^{16}{\rm
O}$+valance-neutron models will be efficient. In the present
calculation, we did not find other state which has the apparent
$\alpha$+$^{16}{\rm O}$+valance-neutrons-like density distribution. The
overlap of the proton wave function with that of $^{20}{\rm Ne}$ wave
function is largest in this state, and 
it decreases in $^{28}{\rm Ne}$ and $^{26}{\rm Ne}$ as neutron number
decreases. The existence of the $0p0h$, $2p2h$ and $4p4h$ 
structure within small excitation energy (they lie within about 5MeV
measured from the lowest  $2p2h$ structure) means the softness of the
neutrons in the $0d_{3/2}$ orbit of this nucleus. Namely, it does not
cost much energy to promote neutrons from the shell which originates in
the spherical $0d_{3/2}$ shell into the higher shell which originates in
the spherical $pf$-shell. It can be attributed to the large
deformation caused by the neutron excitation, the restoration of the
rotational symmetry and the reduced N=20 energy gap owes to the excess of
neutron.
\begin{figure}
\epsfxsize =\hsize
\epsfbox{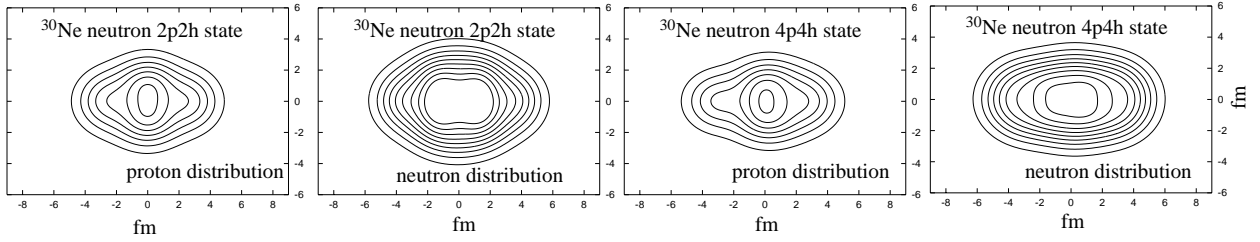}
\caption{Proton and neutron intrinsic density distributions of the
 neutron $4p4h$ structure of $^{30}{\rm Ne}$. For the sake of the
 comparison, those of the $2p2h$ structure are also shown.}
\label{fig::density}
\end{figure}
 
The energy curve of the positive-parity state of $^{28}{\rm Ne}$
(figure \ref{fig::positive_surface} (b)) has the oblately deformed
minimum at $\beta$=0.25 which has a $0p2h$ structure.  We also find the
prolately deformed $2p4h$ structure has a minimum at $\beta$=0.5
whose energy is about 1MeV above the oblately deformed $0p2h$ state.  
This small excitation energy of $2p4h$ structure means that the softness
of neutron's $0d_{3/2}$ orbit again. Since two different structure
coexist within the small excitation energy, when we superpose these wave 
functions, it induces the strong mixing between these configurations
which will be discussed in the next subsection. 
The change of the shape from oblate to prolate is the reason why the
single particle energies as functions of $\beta$ are discontinuous
around $\beta$=0.35 in $^{28}{\rm Ne}$.  The energy of the neutron
$0d_{3/2}$ orbit of   $^{28}{\rm Ne}$ becomes lower as deformation
becomes larger in the oblately deformed region ($0\leq\beta\leq0.45$),
while that of $^{30}{\rm Ne}$ (lower $0d_{3/2}$ of $^{30}{\rm Ne}$)
which is prolately deformed becomes higher in the region of
$0.1\leq\beta\leq0.45$. These behaviors of the  lower $0d_{3/2}$ orbits
are the qualitatively same as those described by the Nilsson
model\cite{ref::NILSSON}, $[N, n_z, m_l,  \Omega]$=[2,1,1,3/2] orbit for
the oblate state and $[N, n_z, m_l, \Omega]$=[2,0,0,1/2] orbit for the
prolate state. Since the neutron number N=18 prefers the oblate
deformation in the small deformed region because of the deformed shell
effect, $0p2h$ structure of $^{28}{\rm Ne}$ has a oblate shape. On the
contrary, the neutron $2p4h$ state has a prolate shape because the
proton number N=10 prefers the prolate deformation, though the neutron's
shell effects are comparable in the prolate and oblate deformation.
In this nucleus, the $4p6h$ structure does not appear as an energy
minimum or a shoulder on the energy curve. When we calculate more deformed
state, it appears about 20MeV above the $0p2h$ state and it does not
seem to be a stable structure.  We think that the absence of the
stable $4p6h$ structure means the hardness of the N=16 deformed shell.
Here, by the deformed N=16 shell, we mean a shell of $sd$ orbits without
the orbits coming from the spherical $d_{3/2}$ orbit.

The energy curve of the positive-parity state of $^{26}{\rm Ne}$ (figure  
\ref{fig::positive_surface} (a)) has an energy minimum  at $\beta$=0.25
which has the $0p4h$ structure. It also has the $2p6h$ shoulder in which
two neutrons are promoted into $pf$-shell, but 
its excitation energy is rather high (about 7MeV in the case of $0^+$
state). We note that the $2p6h$ structure of this nucleus is
superdeformed state. The deformed orbit configuration of this state is
the same as that of the closed shell state with the magic number of the
superdeformation N=16. However the single-particle energy diagram of
figure \ref{fig::positive_spe} (a) does not show clearly the deformed
magic number N=16. 

Here we make some additional comments on the single
particle energy diagrams of Fig.2. Except the  oblate region in
$^{28}{\rm Ne}$, the single particle energy diagrams of the three
isotopes are very similar to each other. From this similarity we can
understand why the deformation of the $2\hbar\omega$-jump structure
becomes larger for lighter Ne isotope reaching the superdeformation in
$^{26}{\rm Ne}$. It is because the uppermost occupied orbit which
crosses first the intruder $pf$-orbit changes to lower $sd$-orbit as
going to lighter isotope, which results in larger value of $\beta$ of
the crossing point.

\begin{figure}
\epsfxsize =\hsize
\epsfbox{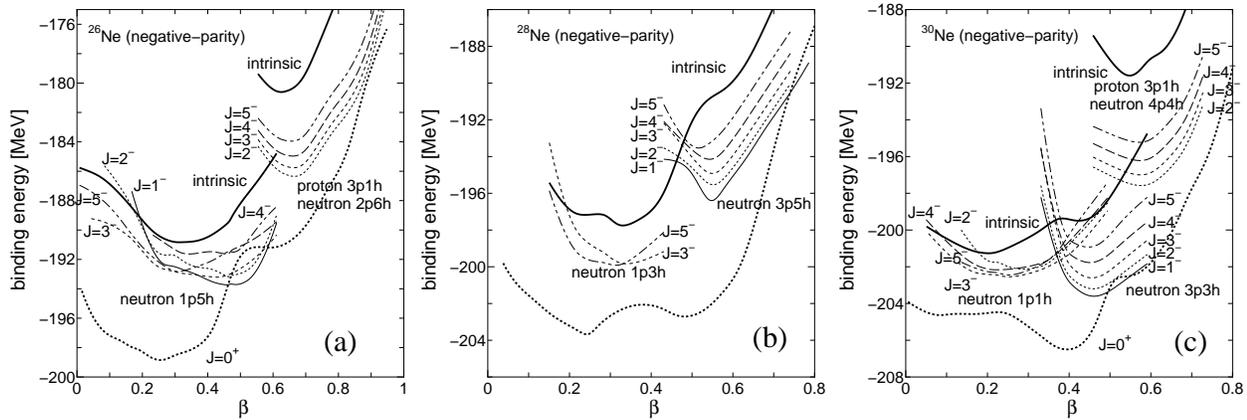}
\caption{Obtained energy curves of the negative parity states of
 $^{26}{\rm Ne}$ (a), $^{28}{\rm Ne}$ (b) and $^{30}{\rm Ne}$ (c). In
 each panel, the energies of the parity projected intrinsic state
 (dashed line), the parity and angular momentum projected states are
 plotted as the functions of the matter quadrupole deformation parameter
 $\beta$.}\label{fig::negative_surface} 
\end{figure}
\begin{figure}
\epsfxsize =\hsize
\epsfbox{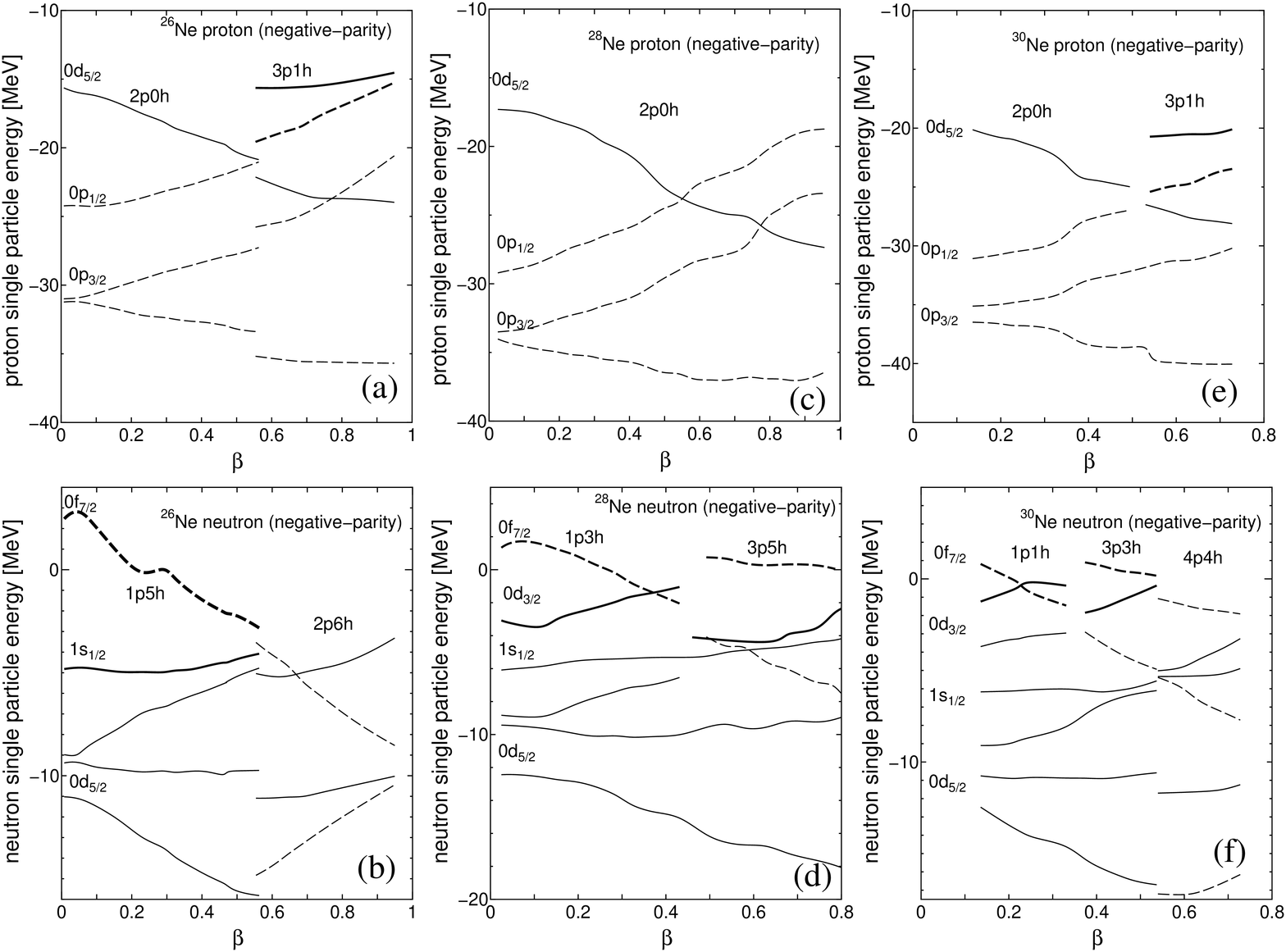}
\caption{The proton and neutron single particle energies of $^{26}{\rm Ne}$ (a) (b),  $^{28}{\rm Ne}$ (c) (d) and $^{30}{\rm Ne}$ (e) (f) obtained from the  intrinsic wave functions on the negative-parity energy curves. The  positive (negative) party single particle orbits are plotted by the  solid (dashed) lines.
The single particle orbits given by the thin (bold) lines are   occupied
 by two (one) protons or neutrons. }\label{fig::negative_spe} 
\end{figure}

Next, we discuss the energy curves of the negative-parity states (figure
\ref{fig::negative_surface}). The energy curves of the positive-parity
states have shown the softness to excite neutrons in the shell which
originates in the spherical $0d_{3/2}$ shell and the hardness to
excite neutrons below N=16. We find that this feature is inherited in
the negative-parity states. Namely, on the one hand the $1p1h$ and
$3p3h$ structure of $^{30}{\rm Ne}$ and $1p3h$ structure of
$^{28}{\rm Ne}$ have small excitation energies.
But on the other hand the $3p5h$ structure of $^{28}{\rm Ne}$
and $1p5h$ structure of $^{26}{\rm Ne}$ have larger excitation
energies. In the 
negative-parity states of $^{30}{\rm Ne}$, there are two low-lying
groups of minimums. The curves which are minimized around $\beta$=0.45 are
the $K^\pi$=$1^-$ rotational band members which have the $3p3h$
structure. Around $\beta$=0.3, there are $J^\pi$=$2^-$, $3^-$, $4^-$ and
$5^-$ states which have the $1p1h$ structure. The order of these
states are changed by the GCM calculation and will be discussed in the
next subsection. The interesting point is that both the $1p1h$
and $3p3h$ structure have small excitation energy, 3.6MeV and
4MeV measured from the positive-parity $2p2h$ state. When we 
compare the energy curves of the positive- and negative-parity states of 
$^{30}{\rm Ne}$, we find that the neutron excitation and nuclear
deformation are correlated to each other. As deformation becomes larger,
neutrons in the shell which originates in the spherical $0d_{3/2}$ shell
are promoted into $pf$-shell, with m=0, 1, 2, 3 and 4, in order for
$mpmh$. The energy  loss due to the neutron promotion is not large and
it always comparable with the energy gain due to the deformation of the
intrinsic state and the restoration of the rotational symmetry. Thus the
neutron 
0-4$ph$ structure appears in the low-lying state. It is
interesting that the $J^\pi$=$1^-$  ($K^\pi$=$1^-$) state which has a
neutron $3p3h$ structure is even lower than the states which has a
neutron $1p1h$ structure. In the shell model calculation, $1p1h$
structure is lower than the $3p3h$ structure\cite{ref:brown}. Though
the energy loss due to the neutron promotion in the $3p3h$ structure is 
larger than the $1p1h$ structure, larger deformation and smaller angular
momentum provide the larger energy gain due to the deformation of the
intrinsic state and the restoration of the rotational symmetry. We also
find an interesting structure around $\beta$=0.6 which appears as a
$K^\pi$=$2^-$ band. Since deformation is too large, neutron $3p3h$
structure seems to be not able to form a stable mean-field. But neutron
$5p5h$ structure costs much energy because of the hardness of the
N=16. Therefore, a proton in $0p$-shell is drafted to make a negative
parity state. Namely, in this structure, the system has a proton $3p1h$
and neutron $4p4h$ structure in which $^{16}{\rm O}$ core is excited. In
other words, this state is understood as the state in which the neutron
$4p4h$ state of the positive-parity state is excited by the proton
excitation. 

The energy curve of the negative-parity state of $^{28}{\rm Ne}$ has two
minima. The lower one which is minimized around $\beta$=0.3 has a
 $1p3h$ structure. It produces $J$=$3^-$ and $5^-$ states and
both have small excitation energy. On the contrary, the $3p5h$
state which produces $K^\pi$=$1^-$ band around $\beta$=0.55 has much
larger excitation energy due to  the hardness of the deformed N=16
shell. Because of the hardness of the N=16 shell closer, the
negative-parity states of $^{26}{\rm Ne}$ do not appear no longer in
the low-energy region. It has $1p5h$ structure around
$\beta$=0.5, but its energy is about 6MeV higher than the
positive-parity $0p4h$ state.
The proton excitation is also found in this nucleus. In the
largely deformed region ($\beta>$0.6), the system chooses the proton
$3p1h$ and neutron $2p6h$ structure instead of the neutron $3p7h$
structure. This choice is because of the stability of the
superdeformed configuration of neutron.

\subsection{Low-Lying Level Scheme}
\begin{figure}
\epsfxsize =0.9\hsize
\epsfbox{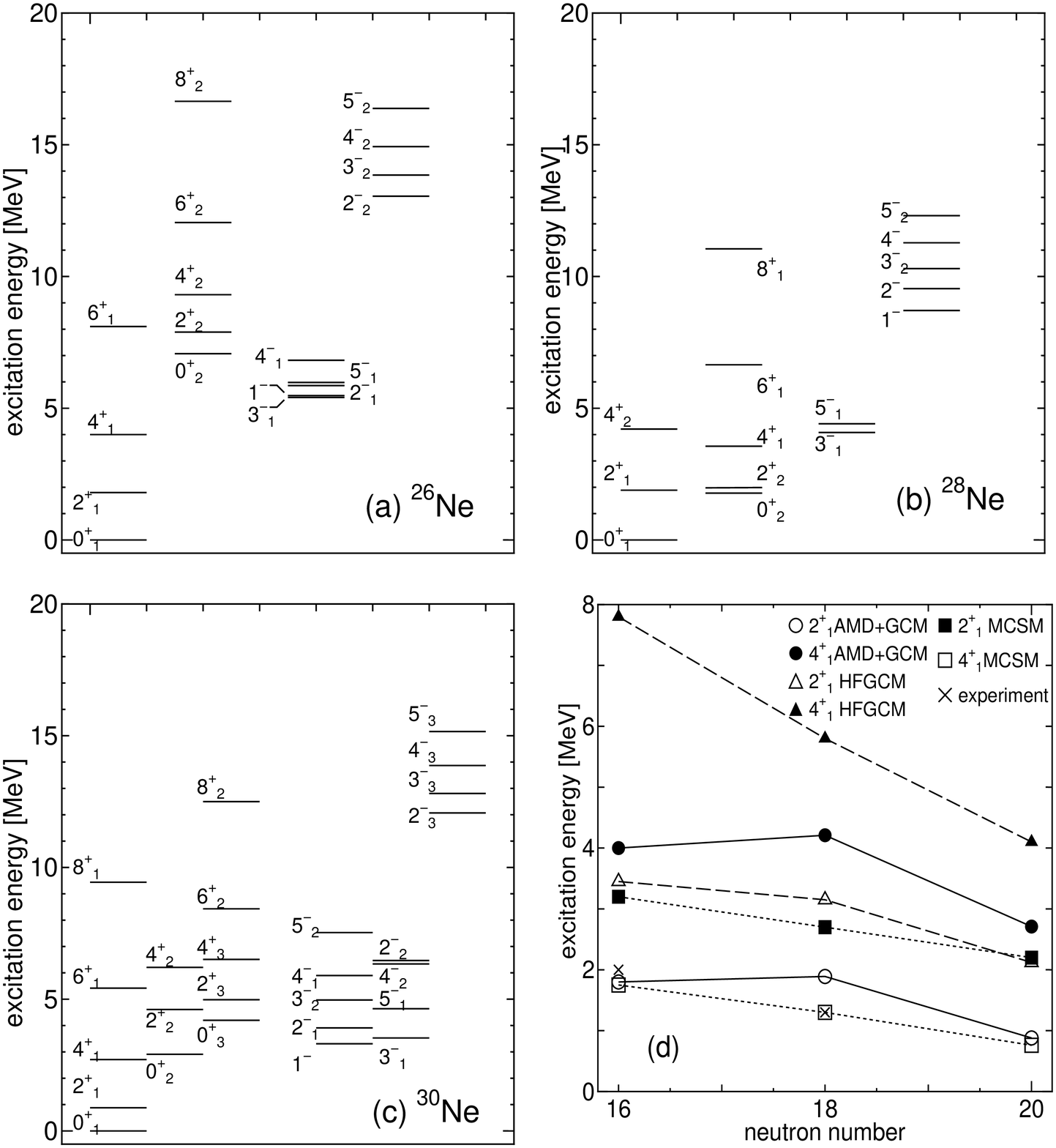}
\caption{Calculated low-lying level schemes of  (a) $^{26}{\rm Ne}$,
 (b) $^{28}{\rm Ne}$ and  (c) $^{30}{\rm Ne}$. (d) shows the comparison
 of the excitation energies of  $2^+_1$ and $4^+_1$ states obtained by
 the present work, the monte carlo shell model\cite{ref:utsuno_1} (MCSM),
 HFB+AMPGCM\cite{ref:egido_2} and the experiment.}\label{fig::levels}
\end{figure}

\begin{wraptable}{r}{\halftext}
\caption{$E2$ transition probabilities. Experimental data
 is taken from Ref \cite{ref:pritychenko}.} \label{tab::BE} 
\begin{tabular}{r|l|c}
& AMD+GCM&  EXP  \\\hline 
$^{30}{\rm Ne}$; $0^+_1 \rightarrow 2^+_1$& 283 & \\
$0^+_2 \rightarrow 2^+_2$& 88  &\\ 
$0^+_3 \rightarrow 2^+_3$& 393 &\\\hline 
$^{28}{\rm Ne}$; $0^+_1 \rightarrow 2^+_1$& 208 &269$\pm$136 \\ 
$0^+_1 \rightarrow 2^+_2$& 112 &\\
$0^+_2 \rightarrow 2^+_1$& 109 &\\
$0^+_2 \rightarrow 2^+_2$& 239 &\\\hline
$^{26}{\rm Ne}$; $0^+_1 \rightarrow 2^+_1$ & 203  & 228$\pm$41 \\
$0^+_2 \rightarrow 2^+_2$& 412 & 
\end{tabular}
\end{wraptable}

After the angular momentum projection, we have superposed the wave
functions calculated by the variation and diagonalized (GCM
calculation). The obtained level scheme of $^{26}{\rm Ne}$, $^{28}{\rm
Ne}$ and $^{30}{\rm Ne}$ is shown figure \ref{fig::levels}. First, we
discuss $^{30}{\rm Ne}$. In the positive-parity state, we have obtained
three bands. The ground, second and third bands have dominant
contribution from $2p2h$, $0p0h$ and $4p4h$ structure, respectively. The
mixing between these three neutron configurations is most strong in the
$0^+$ state and becomes smaller as the angular momentum increases.
Similarly to the results of other theoretical studies, $2^+_1$
state has a small excitation energy (0.88 MeV) and large 
B($E2$;$0^+_1\rightarrow2^+_1$) value (table \ref{tab::BE}) due to the
dominance of the large deformed $2p2h$ configuration in the ground
band. The mixings between different neutron  
configurations are weaker in the $2^+_1$ and $4^+_1$ states than in the
ground state. Therefore the mixing between the neutron configurations
lowers the ground state energy than the higher spin 
states and makes the 
spectrum of the ground band slightly deviate from the rotational one,
$E(2^+_1)/E(0^+_1)$=3.08.  The second band shows the vibrational 
spectra. The third band has a $4p4h$ dominant structure and this band
shows the largely deformed rotational spectrum and has large intra-band
$E2$ transition probabilities (table \ref{tab::BE}). 
Even though we increase the number of the basis states of the GCM
calculation, the energy and wave function of the third band is
not affected as well as those of the ground and second bands,
and therefore, we consider that this band is stable. The small
excitation energy of this band is important, since it means the softness
of the neutrons above the deformed N=16 shell. In the same sense, the
low-lying negative-parity states are also important, though their
spectra are complicated due to a rather strong mixing between $1p1h$ and
$3p3h$ structures. As discussed in the previous section, due to the
large energy gain which owe to the larger deformation and the smaller
angular momentum, $1^-$ state becomes the lowest negative-parity
state. $3^-_1$ state is lowered by the strong mixing between $1p1h$
states with the $3p3h$ structure and in this state, $1p1h$ structure is
dominant, while $3^-_2$ state becomes higher and have a $3p3h$ dominant
structure.  $5^-_1$ state is also
lowered by the mixing between two configurations and in this state,
neutron $1p1h$ structure is dominant.  There are not strong mixing
between $2^-_1$ and $2^-_2$ and between $4^-_1$ and $4^-_2$ state and it
leaves $2^-_2$ and $4^-_2$ states which have $1p1h$ dominant structure
in higher excitation energy than the $5^-_1$ and $3^-_1$ states which
also have $1p1h$ structure. The $K^\pi$=$2^-$ band in which a proton in
$p$-shell is excited into $sd$-shell combined with the neutron $4p4h$
structure  does not mixed with other states and left in the higher
energy region. The obtained low-lying states of $^{30}{\rm Ne}$ have been
governed by the neutron's particle-hole configurations. The softness of 
neutrons in the $0d_{3/2}$ orbit against the promotion into the
$pf$-shells produces the low-lying $0p0h$ and $2p2h$ rotational
bands in $^{30}{\rm Ne}$. This trend is also common in the case of the
negative-parity states. In particular, the lowest negative-parity band,
$K^\pi$=$1^-$ band which has neutron $3p3h$ structure dominantly, is
as largely deformed as the ground band. And it can be regarded as the
$1\hbar\omega$  excitation from the ground band in the same deformed
mean-field as the ground band. The small excitation energy
$K^\pi$=$0^+_3$ band which has neutron $4p4h$ structure dominantly can
be also regarded as showing the softness of the $0d_{3/2}$
orbit. However, as is already presented, this band has a
`$\alpha$+$^{16}{\rm O}$+valance neutrons' type nature like the case of
the neutron-rich Be isotopes\cite{Enyo,Itagakio}. It means the
coexistence of the deformed mean-field structure and the molecular
orbital structure and their interplay which has not clearly seen in the
case of the Be isotopes.

The ground band of $^{28}{\rm Ne}$ has a strong mixing between the
oblately deformed $0p2h$ structure and the prolately deformed
$2p4h$ structure. The mixing is strongest in the case of the $0^+$ states
and it lowers the ground state's energy  by about 1.9MeV. This mixing
becomes smaller rapidly as the angular momentum 
becomes larger. As a result, the energy gap between $0^+_1$ state and
$2^+_1$ state becomes larger and the $0^+_1$, $2^+_1$ and $4^+_1$
states of the ground band show vibrational character,
$E_x(4^+_1)$/$E_x(2^+_1)$=2.17. The strong mixing also makes the $E2$
transition probability between the ground state and $2^+_1$ state
smaller compared to that of $^{30}{\rm Ne}$ (table \ref{tab::BE}), since
the amount of the largely  deformed component becomes smaller than the
ground band of $^{30}{\rm Ne}$.
The excitation energy of $0^+_2$ becomes higher by the mixing. And this
is the reason of the small energy gap between $0^+_2$ and 
$2^+_2$ states. In the case of the $^{30}{\rm Ne}$, the low-lying
$1p1h$ and $3p3h$ states are mixed strongly. However, in this nucleus,
the mixing between $1p3h$ and $3p5h$ structure is no longer strong
because of their large energy difference. As a
result, the $3^-_1$ and $5^-_1$ states which have $1p3h$ structure
appears in the lower energy region together with the positive-parity
bands, while $K^\pi$=$1^-$ band which has $3p5h$ structure is left in
the higher energy region. The softness of neutrons in $0d_{3/2}$ induces
the strong mixing between neutron $0p2h$ and $2p4h$ configurations, but
the hardness of deformed neutron N=16 shell makes the mixing between
$1p3h$ and $3p5h$ state weak.

In the ground band of the $^{26}{\rm Ne}$, the mixing between the
different neutron configurations is much weaker than $^{30}{\rm Ne}$ and 
$^{28}{\rm Ne}$. The ground band of the $^{26}{\rm Ne}$ shows a
vibrational character, $E_x(4^+_1)/E_x(2^+)$=2.22. Due to the hardness
of the N=16, neutron $ph$ states appears more than 5MeV above the ground
state. In the positive-parity the superdeformed states which has $2p6h$ 
structure appears about 6MeV above the ground state. It seems that the
magic number of the superdeformation N=16 which is most detailedly
studied in $^{32}{\rm S}$ is still valid even in the neutron-rich
region. Indeed, in this superdeformed state, it is possible to excite a
proton from $p$-shell to $sd$-shell by not affecting the stability of
this state. This proton excited states appears as the $K^\pi$=$2^-$
state in the negative-parity. The lowest negative-parity state is
$3^-_1$ state but its excitation energy is higher than that of the
$^{30}{\rm Ne}$ and  $^{28}{\rm Ne}$. Because of the hardness of the
deformed neutron N=16 shell, this nucleus does not have any low-lying
particle-hole excited states. Instead, it has a superdeformed rotational
band about 7MeV above the ground band whose neutron configuration is 
almost the same as that of the predicted superdeformed state of
$^{32}{\rm S}$\cite{yamagami,kimura}. This superdeformed neutron
configuration seems to be stable, since it allows the proton excitation
within the same superdeformed mean-field which appears as the
$K^\pi$=$2^-$ band about 13MeV above the ground state.

\begin{wrapfigure}{t}{\halftext}
\epsfxsize =\hsize
\epsfbox{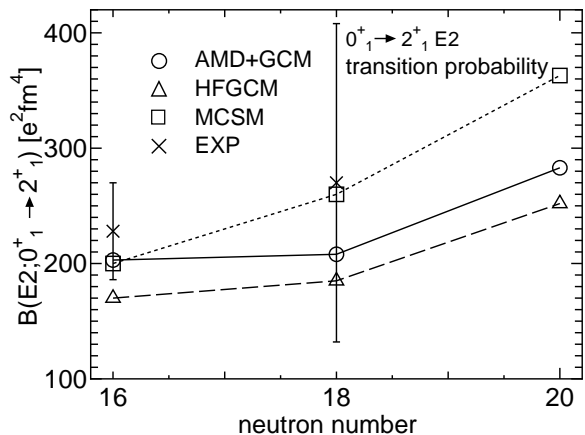}
\caption{Calculated and observed B($E2$;$0^+_1\rightarrow2^+_1$) of Ne
 isotopes.   For comparison, results by monte carlo shell model
 (MCSM)\cite{ref:utsuno_1} and  HFB+AMPGCM\cite{ref:egido_2} are
 shown. }\label{fig::E2} 
\end{wrapfigure}

Finally, we discuss the comparison with the experimental data and the
other theoretical results. The excitation energy of $2^+_1$ state of
$^{26}{\rm Ne}$ shows a 
good agreement with the experiment. However, in our result, the observed
$0^+_2$ state is not obtained. We note that the experimentally
observed $0^+_2$ state at 3.69MeV will not correspond to the $0^+_2$
state obtained by the present calculation. In the HFB+GCM
calculation\cite{ref:egido_2}, the prolately deformed local minimum which
may have a neutron $0p4h$ structure appears and that will be correspond
to the experimentally observed $0^+_2$ state. Since in the variational
calculation, we have made the constraint on the quadrupole deformation
parameter $\beta$, but not on the $\gamma$, we have not obtained the
prolately deformed minimum which have neutron $0p4h$ structure on the
energy curve. The excitation energies of $2^+_1$ and $4^+_1$ states of
$^{28}{\rm Ne}$ and $^{30}{\rm Ne}$ is smaller than those obtained by
HFB+GCM calculation\cite{ref:egido_2} and larger than the monte carlo
shell model study\cite{ref:utsuno_1} (figure \ref{fig::levels} (d)),
though the general trend of these state is common. The difference
between our results and HFB+GCM results may be attributed to the
difference of the spatial symmetry of the wave function, the difference
of the  generator coordinates used in the calculation and the lack of
the paring effects in our calculation. However, in the
B($E2$;$0^+_1\rightarrow 2^+_1$) we do not find such large difference 
between our results and HFB+GCM results. In $^{32}{\rm Mg}$, the $3^-$ 
state at 2.32 MeV is  experimentally
proposed\cite{ref:Klotz,ref:mittig}, though the assignment of the spin
and parity of this state is not still  fixed. This state can correspond
to the $3^-_1$ state of $^{30}{\rm Ne}$ at 3.31MeV obtained in the
present calculation.    

\section{Summary}
We have studied the low-lying level schemes of $^{26}{\rm Ne}$,
$^{28}{\rm Ne}$ and $^{30}{\rm Ne}$ using the deformed-basis AMD+GCM
method. The obtained energy curves and the level schemes of the positive
parity states have shown the softness of the neutrons in the deformed
orbits coming from the $0d_{3/2}$
orbit with respect to the promotion to the $pf$-shell and the hardness
of neutrons below N=16 deformed shell. The softness leads to the
coexistence of the 
neutron $0p0h$, $2p2h$ and $4p4h$ states in $^{30}{\rm Ne}$ and $0p2h$
and $2p4h$ states in $^{28}{\rm Ne}$ and to the breaking of the N=20
shell closure in $^{30}{\rm Ne}$. We also found that this feature is
inherited to the negative-parity states As a result, neutron $1p1h$ and
$3p3h$ states of $^{30}{\rm Ne}$ and $1p3h$ state of $^{28}{\rm Ne}$ are
predicted to appear in the small excitation energy. We think that the
low excitation energies of negative parity states are typical phenomena
accompanying the breaking of N=20 shell, but it has not been
recognized because almost no theoretical studies including Hartree-Fock
approach have been made an negative parity states. The proton excited
states in the negative parity states of $^{30}{\rm Ne}$ and $^{26}{\rm
Ne}$ which are the results of the hardness of the N=16 deformed shell have
been also obtained. The $\alpha$+$^{16}{\rm O}$ cluster correlations
have been noticed to exist in the $4p4h$ structure of $^{30}{\rm Ne}$.

\section*{Acknowledgements}
We would like to thank Dr. Y. Kanada-En'yo for valuable discussions.    
Many of the computational calculations were carried out by 
SX-5 at Research Center for Nuclear Physics, Osaka University (RCNP). 
This work was partially performed in the Research Project for Study of
Unstable Nuclei from Nuclear Cluster Aspects¡É sponsored by Institute of
Physical and Chemical Research (RIKEN).

\end{document}